\documentstyle[aps,prl]{revtex}
\begin{document}
\input epsf.sty
\twocolumn[\hsize\textwidth\columnwidth\hsize\csname %
@twocolumnfalse\endcsname
\draft
\widetext

\title{X-ray-induced disordering of the dimerization pattern and apparent
low-temperature enhancement of lattice symmetry in 
spinel CuIr$_2$S$_4$}

\author{H. Ishibashi$^{1,2}$, T. Y. Koo$^1$, Y. S. Hor$^1$, A.
Borissov$^1$, Y. Horibe$^{3,4}$, P. G. Radaelli$^{5,6}$,
S-W. Cheong$^{1,3}$, and V. Kiryukhin$^1$}
\address{(1) Department of Physics and Astronomy, Rutgers University,
Piscataway, New Jersey 08854}
\address{(2) Department of Materials Science, Osaka Prefecture University,
Sakai, Osaka 599-8531, Japan}
\address{(3) Bell Laboratories, Lucent Technologies, Murray Hill,
New Jersey 07974}
\address{(4) Department of Materials Science and Engineering and Kagami
Memorial Laboratory for Materials Science and Technology, Waseda University,
Shinjuku-ku, Tokyo 169, Japan}
\address{(5) ISIS Facility, Rutherford Appleton Laboratory, Chilton,
Didcot, Oxfordshire, OX11 0QX, United Kingdom}
\address{(6) Dept. of Physics and Astronomy, University College London,
Gower St., London, WC1E 6BT, United Kingdom}

\date{\today}
\maketitle

\begin{abstract}

At low temperatures, spinel CuIr$_2$S$_4$ 
is a charge-ordered spin-dimerized insulator with triclinic lattice symmetry.
We find that x-rays induce a
structural transition in which the local triclinic structure is preserved,
but the average lattice symmetry becomes 
tetragonal. These structural changes are accompanied by a thousandfold
reduction in the electrical resistivity.
The transition is persistent, but
the original state can be restored by thermal annealing.
We argue that x-ray irradiation disorders
the lattice dimerization pattern, producing a state in which
the orientation of the dimers is preserved, but the translational long-range
order is destroyed.  

\end{abstract}


\phantom{.}
]
\narrowtext

Spinel compounds $AB_2X_4$ ($X$ is O, S, Se, or Cl)
have attracted much attention over the last decade
because they exhibit a large variety of interesting ground states,
including superconductivity, cooperative antiferromagnetism, 
heavy fermion, and charge-ordered
and spin-dimerized states \cite{Review,Spinels}. The panoply of different
properties exhibited by spinels results from the
interplay of Coulomb interactions, effects of frustrated magnetism, and 
electron-lattice interaction. The corner sharing tetrahedral network of $B$
sites in the spinel structure can accommodate a large number of different charge
ordering patterns. In fact, some of the most complex charge-ordering patterns
reported to date are found in spinel compounds \cite{Radaelli,Li}. 
The same tetrahedral network
gives rise to geometric magnetic frustration when the ions occupying the $B$
site are magnetic. The electronic and magnetic states realized in such a
complex environment are often multi-degenerate and strongly
fluctuating. Because of these complexity, a number of properties of the 
spinels remain poorly understood. Spinels, therefore, are important subjects
of research in the physics of strongly correlated materials. In addition, 
some of these compounds,
such as the lithium manganese spinels used in battery cathodes and the ferrites
used in microwave applications, are of
substantial technological importance. 

Chalcogenide spinel
CuIr$_2$S$_4$ has attracted attention because this compound undergoes a sharp
metal-insulator transition at T$_{MI}\approx$230 K \cite{MIT}.
The low-temperature insulating state is nonmagnetic.
NMR and photoemission experiments have shown that the Cu
ion is monovalent in the insulating phase, and therefore the nominal
valence of the iridium atoms is 3.5 \cite{CuIon}. 
It was proposed that charge ordering of Ir$^{3+}$ (S=0) and 
Ir$^{4+}$ (S=1/2) ions is the origin of the metal-insulator
transition in CuIr$_2$S$_4$ \cite{MIT}, and that some kind of spin dimerization
is responsible for the nonmagnetic nature of the insulating phase. 

\begin{figure}
\centerline{\epsfxsize=2.9in\epsfbox{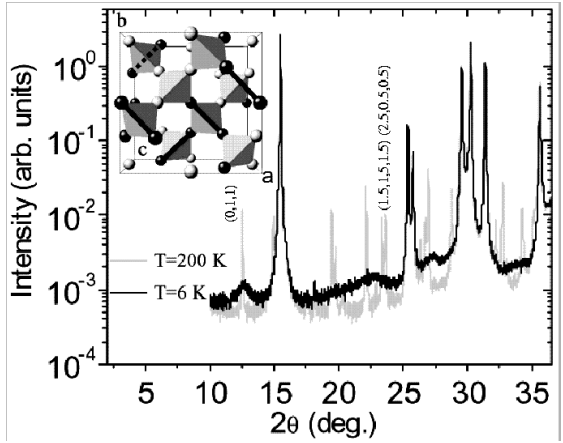}}
\vskip 5mm
\caption{X-ray powder scans at T=200 K and T=6K, $\lambda$=1.5406$\rm\AA$.
The inset shows the
charge ordering pattern in the triclinic state. Ir$^{3+}$ and Ir$^{4+}$ ions
are shown in white and black, respectively. Black bonds indicate dimerized
Ir$^{4+}$ ions.}
\label{fig1}
\end{figure}

Recent 
experimental determination of the low-temperature structure \cite{Radaelli}
has provided very strong evidence that this scenario is correct. Specifically,
these experiments have shown that the Ir sublattice consists of two types of
Ir bi-capped hexagonal rings, which were described as Ir$^{3+}_8$ and 
Ir$^{4+}_8$ octamers. The S=1/2 Ir$^{4+}$ ions were found to form structural
dimers, which were also identified as spin dimers. 
The magnitude of the lattice distortion due to this dimerization is
truly remarkable: the Ir-Ir distance in the dimers is $\sim$3.0$\rm\AA$, while
all the other Ir-Ir nearest neighbor distances are between 3.43 and 3.66
$\rm\AA$. Inset in Fig. \ref{fig1} illustrates the low-temperature charge
ordering and spin-dimerization pattern reported in \cite{Radaelli}.  
The low-temperature state possesses triclinic symmetry \cite{Hiroki,Radaelli}.
However, for simplicity, we describe the structure of CuIr$_2$S$_4$ using the
conventional cubic spinel unit cell, which is not the true unit cell at low
temperatures.

In this work, we report high-resolution x-ray powder diffraction
and simultaneous electrical resistance measurements on CuIr$_2$S$_4$. 
We find that at low temperatures, x-rays induce a
triclinic-to-tetragonal structural transition in which the electrical
resistivity is reduced by a factor of 10$^3$. The transition is persistent, but
the original state can be restored by heating above T$\sim$100 K and
subsequent cooling.
To our knowledge, the only other example of such a dramatic reversible
x-ray-induced structural change is found in perovskite manganites, in which
x-rays convert a charge ordered insulator to ferromagnetic metal \cite{Mang}.
In our samples,
all the Bragg peaks in the x-ray converted state can be accounted for using the
tetragonal $I4_1/amd$ space group. However,
analysis of the diffuse x-ray scattering in the tetragonal state
shows that the triclinic structure is preserved locally.
Therefore, x-ray irradiation changes only the average (or global) symmetry.
We argue that this apparent change in the 
lattice symmetry is caused
by x-ray-induced disordering of the lattice dimerization pattern, in which
the orientation of the Ir$^{4+}$ dimers is preserved. Tetragonal
CuIr$_2$S$_4$ provides, therefore, an interesting example of a 
state possessing rotational, but not translational, long range dimer order.

Polycrystalline samples of CuIr$_2$S$_4$ were prepared by a solid reaction 
method. The mixtures of Cu, Ir, and S powders were heated in an evacuated
quartz tube at 800-900 $^\circ$C
for 8 days. The samples were then ground, pressed
into a pellet, and sintered in vacuum at 900 $^\circ$C for another 3 days.
X-ray powder
diffraction measurements were carried out at beamlines X20C and X25
at the National
Synchrotron Light Source. An x-ray beam was focused by a mirror, monochromatized
by a double-crystal Si (111) monochromator, and analyzed with Ge (111) crystal. 
The wavelengths used were $\lambda$=1.513 and 1.5406
$\rm\AA$, the beam size was $\sim$1 mm$^2$, and the typical intensity was
3$\cdot$10$^{11}$ photons/sec. The sample was mounted
in a closed-cycle refrigerator (T=6-300 K). The powder patterns were measured
using the
$\theta/2\theta$ step scanning mode.
For electrical resistivity measurements, four gold contacts were evaporated
on the surface of the polycrystalline sample. The 
x-ray beam was hitting the sample
in between the contacts, and x-ray diffraction and electrical resistance
measurements were done simultaneously. Neutron powder diffraction data were
collected using the HRPD high-resolution powder diffractometer
at the ISIS facility.

Fig. \ref{fig1} shows partial x-ray scans taken at T=6 K and T=200 K. 
Note that the x-ray
intensity is shown on a logarithmic scale. The T=6 K scan was taken after 
several hours of initial exposure to x-rays. All the Bragg peaks at T=6 K can
be indexed assuming the tetragonal symmetry with the space group $I4_1/amd$ and
lattice constants $a$=6.8766 $\rm\AA$ and $c$=10.039 $\rm\AA$, with the $a$ and $b$ axes pointing along the (110) and (1-10) directions in the cubic spinel
unit cell. At T=200 K, the structure is triclinic. For T$>$T$_{MI}\approx$230 K,
CuIr$_2$S$_4$ exhibits an undistorted cubic spinel structure.
In this work, all the Bragg
peaks are indexed using the cubic spinel unit cell with
lattice constant $a\sim$9.8$\rm\AA$. Note, this indexing scheme is
only approximate for $T<T_{MI}$. In the
triclinic state, a number of peaks that are not present in the cubic and the
tetragonal phase appear (see Fig. \ref{fig1}). We refer to these peaks as
superlattice peaks. 

\begin{figure}
\centerline{\epsfxsize=2.9in\epsfbox{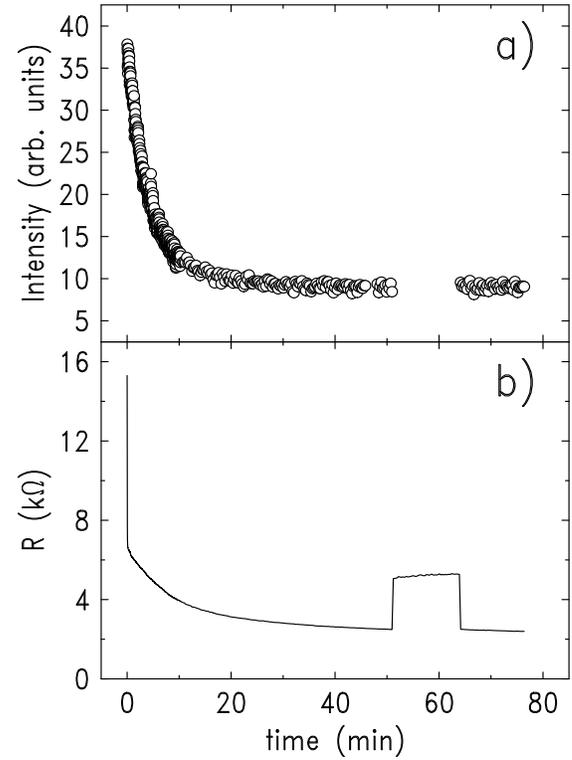}}
\vskip 5mm
\caption{X-ray exposure dependence of (a) the intensity of the
(2,2,1) superlattice
peak and (b) electrical resistance at T=10 K. X-rays were switched off between
$t$=51 min and $t$=64 min.}
\label{fig2}
\end{figure}

According to our neutron diffraction measurements taken after the sample was 
equilibrated for 12 h at T=4 K, the low-temperature state of 
CuIr$_2$S$_4$ is triclinic.
The tetragonal state is induced by x-rays. Figure \ref{fig2}(a) shows that 
a 30 min x-ray exposure completely destroys the superlattice peaks. 
At the same time, the resistance of the sample is significantly reduced,
see Fig \ref{fig2}(b). When the 
x-rays are switched off, the resistance increases
abruptly (most likely due to beam heating effects), 
and then keeps increasing with a time constant $\tau >$5 h.
It is possible, therefore, that the x-ray induced state is metastable.
Even after a 10 h wait, however,
the resistance does not reach a third of its original
value, and no detectable superlattice peaks are observed after 24 h. 
In our experimental geometry, x-rays
penetrated only $\sim$2 $\mu$m into the sample, and therefore only a small
surface layer of the sample was undergoing the x-ray-induced transition.
Assuming that the thickness of this layer equals to the x-ray penetration
depth, we estimate that the resistivity of the tetragonal state is approximately
2 $\Omega$cm. This is at least 
3 orders of magnitude smaller than the resistivity of
the triclinic phase in polycrystalline CuIr$_2$S$_4$ \cite{MIT,Hiroki1}.

\begin{figure}
\centerline{\epsfxsize=2.9in\epsfbox{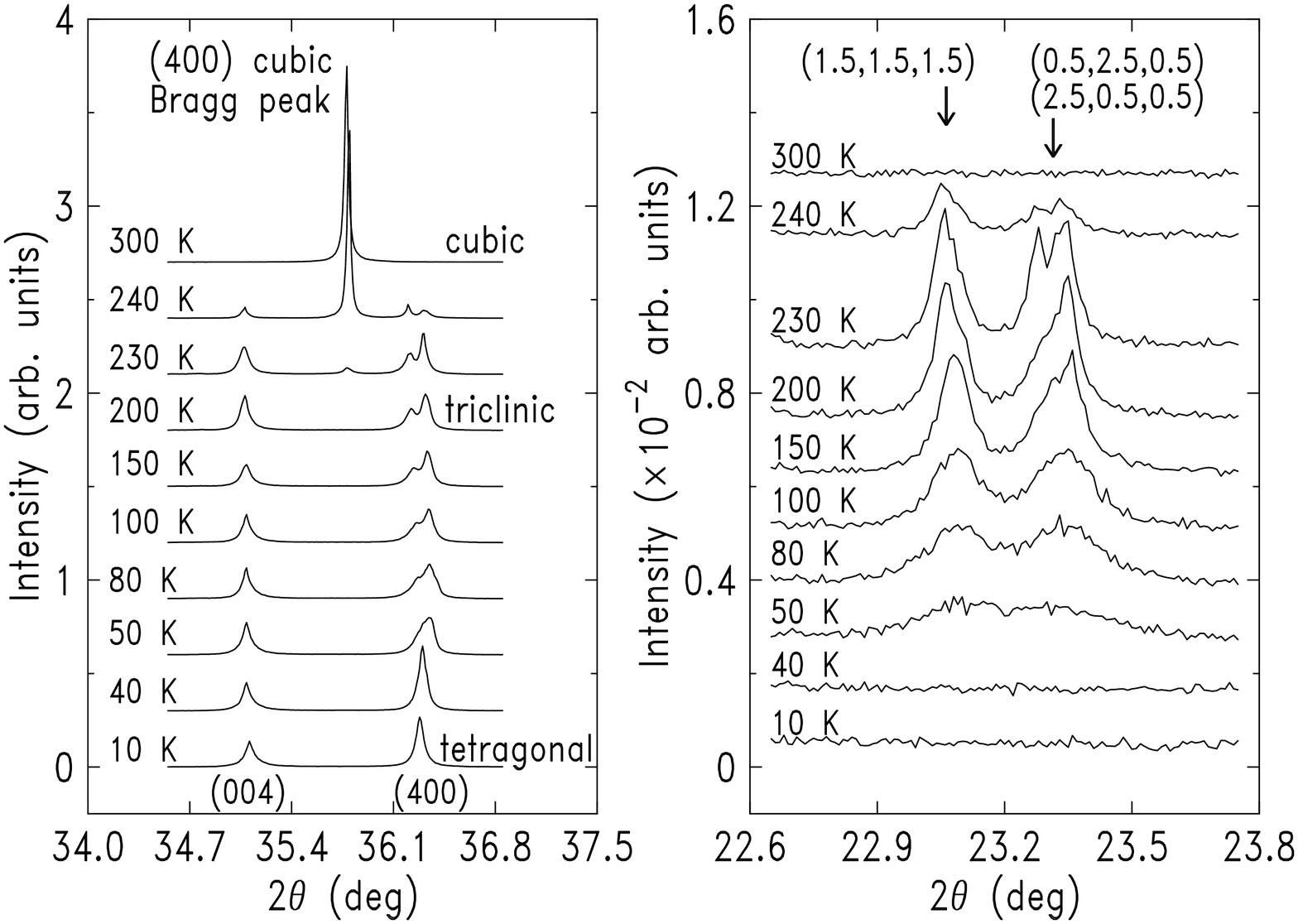}}
\vskip 5mm
\caption{X-ray scans in the vicinity of the (400) cubic Bragg peak, and the
(1.5,1.5,1.5), (0.5,2.5,0.5), and (2.5,0.5,0.5) superlattice peaks at 
various temperatures, taken on heating. X-ray wavelength $\lambda$=1.513
$\rm\AA$.}
\label{fig3}
\end{figure}

The triclinic state recovers when the sample is heated above 100 K, even
in the presence of x-rays. In Figure \ref{fig3}, scans in the vicinity of the
(4,0,0) main Bragg peak and the (1.5,1.5,1.5) superlattice peak taken on
heating at different temperatures are shown. The triclinic state manifests
itself by splitting of the tetragonal peaks and by the appearance of the
superlattice peaks. The superlattice peaks are very broad at low temperatures.
A careful examination of the x-ray scan at T=6 K in Fig. \ref{fig1}
shows that enhanced diffuse scattering exhibiting broad features with maximums
near the superlattice peak positions is present in the tetragonal state.
Such a broad feature can be
seen, for example, at the (0,1,1) superlattice peak position at 
2$\theta$=12.7$^\circ$. Broadening of the diffraction peaks is associated with 
finite correlation length of the ordered state. Therefore,
short-range-ordered triclinic regions, and the associated Ir$^{4+}$ dimers,
are present even in the x-ray converted 
state. Unlike the superlattice peaks, the main Bragg peaks remain narrow
in the entire temperature range.

\begin{figure}
\centerline{\epsfxsize=2.9in\epsfbox{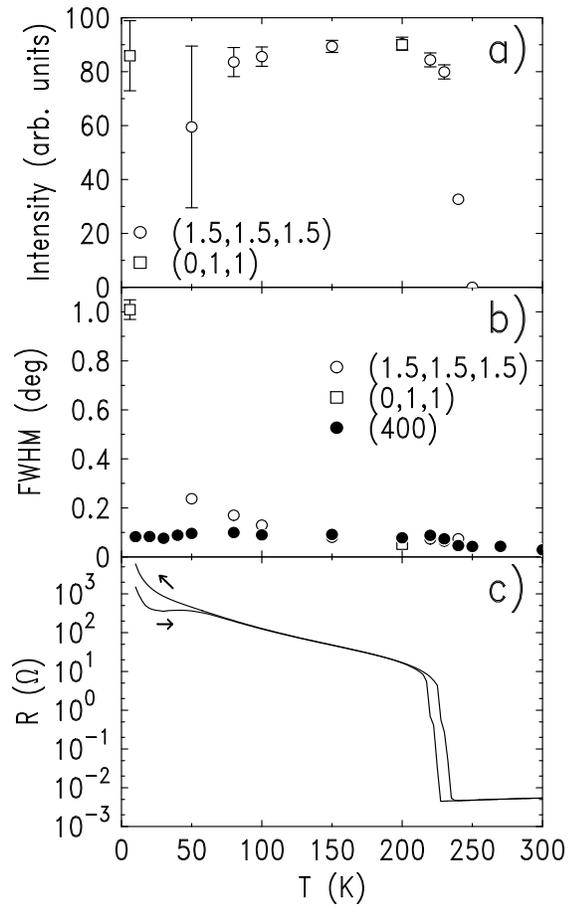}}
\vskip 5mm
\caption{Temperature dependence of (a) the integrated intensity of the
(1.5,1.5,1.5) and (0,1,1) superlattice peaks, (b) full width at
half maximum of the same superlattice peaks and of the (400) Bragg peak, and
(c) electrical resistance. The data were taken on heating from the x-ray
converted state. Electrical resistance taken on cooling with no
x-rays present is also shown. 
The (0,1,1) peak intensity in (a) was normalized to
match the (1.5,1.5,1.5) intensity at T=200 K.}
\label{fig4}
\end{figure}

Figure \ref{fig4} shows the integrated intensity of the (0,1,1) and
(1.5,1.5,1.5) superlattice peaks, and the superlattice peak width taken
at different temperatures on
heating from the x-ray converted state. These data were obtained from
the scans shown in Figs. \ref{fig1}, \ref{fig3} using Lorentzian fits.
The integrated intensity is
the same in both the triclinic and the tetragonal states, within
experimental errors. Therefore,
on the local level, the structure is triclinic everywhere in the x-ray
converted state. The correlation length of the
triclinic lattice distortion, estimated from the width of the (0,1,1)
peak, is $\xi_{tr}\sim$30 $\rm\AA$ at T=6 K.

Dimerization of Ir$^{4+}$ is the dominant
source of the triclinic lattice distortion, and therefore
the long-range correlation between the
dimers is lost in the tetragonal state. 
We propose the following scenario of the x-ray
induced transition. X-rays remove localized electrons from the
Ir atoms, thereby instantaneously destroying local
charge-ordering and lattice dimerization pattern. The electrons quickly relax,
and the charge-ordering is restored. However, Ir$^{4+}$ dimers
do not necessarily form in the original places. This is, in fact, not
unexpected, if one takes
into account that slowly propagating elastic interactions
play a dominant role in the formation of the dimerization pattern in
CuIr$_2$S$_4$. The Coulomb interaction is less important in
this compound, as indicated by the breakdown of the Anderson's condition which
requires that two Ir$^{3+}$ and two Ir$^{4+}$ ions are present in every
Ir tetrahedron (see inset in Fig. \ref{fig1}). 
It is interesting to note that the correlation length $\xi_{tr}$ is large
enough to accommodate several Ir octamers. This indicates 
that the Ir$^{4+}$ dimers still tend to form the octamer units
in the tetragonal state, and therefore these octamers are quite robust.

Rietveld refinement of the powder data in the x-ray converted
tetragonal state can only provide the average structure. In fact, in the
$I4_1/amd$ group, all the Ir atoms are equivalent, and no direct information
about the Ir charge order can be obtained in a conventional refinement. 
Experiments on single 
crystals will eventually be required to determine the  local structure 
of the tetragonal state, and XPS or NEXAFS measurements will be needed to
study the valence state of the Ir atoms in the irradiated samples.
We have, however,
attempted the Rietveld refinement of the powder data at T=6 K putting 
Ir$^{3+}$ ions in the {\it 8d} site with occupation 0.5, and Ir$^{4+}$ ions in
the {\it 16h} site with occupation 0.25. 
The refinement produced two Ir$^{4+}$ ions
in each {\it 16h} site displaced along the [110] cubic direction with respect to
each other. Since in the triclinic state the Ir$^{4+}$ dimers form along the
same direction \cite{Dir}, 
this result is consistent with the assumption that uncorrelated
dimers are present in the sample. It also indicates that the dimers in the
x-ray converted state form along the same direction as in the triclinic phase.
Moreover, the intradimer distance obtained
in our refinement, 3.09 $\rm\AA$, is close to the actual value found in the
triclinic state. While the overall quality of the fit is only average 
($R_{wp}$=0.100), the fit is consistent with the disordering scenario proposed
above. Further details of the Rietveld refinement will be given elsewhere
\cite{Hiroki1}.

In addition to the results of the Rietveld refinement, the increased value
of the $c$ lattice constant in the tetragonal state provides an
evidence that the Ir$^{4+}$ dimers remain contained in the $ab$ plane and,
therefore, do not change their orientation. The long-range translational
order of the dimers, on the other hand, is clearly destroyed in the tetragonal
state. An interesting question, which requires further investigation, is
whether the tetragonal state in CuIr$_2$S$_4$ could be considered a 
thermodynamic phase,
with some properties, possibly, resemling those of liquid crystal phases.
To answer this question, the equilibrium thermal properties
of the tetragonal state need to be investigated. As was mentioned above, 
the x-ray induced state is in all likelihood metastable. However, it might
be possible that the equilibrium tetragonal state can be induced by other
sources of disorder. In particular, static disorder from chemical
substitution appears to be promising.
To test this proposal, experiments with
Cu(Ir,Cr)$_2$S$_4$ samples are currently under way.

The reduced electrical resistivity in the x-ray converted state most likely
results from the imperfections in the charge order. Charge-disordered 
conducting regions might form, for instance, at the boundaries between the
ordered triclinic domains.
Consistent with this assumption, the resistivity
characteristic to the triclinic state is recovered on heating, as these 
imperfections disappear and the long-range
ordered triclinic state is restored, see Fig. \ref{fig4}(c).

Finally, we note that in a recent paper, Sun {\it at al.}
reported an electron diffraction experiment suggesting a structural
transition in CuIr$_2$S$_4$ at T$<\sim$50 K \cite{Sun}. We speculate that this
transition is of the same nature as the x-ray induced transition described 
above, and that it is induced by an electron beam. 

In summary, we find that x-rays induce an apparent triclinic-to-tetragonal
transition in CuIr$_2$S$_4$ at low temperatures. In addition to the drastic
structural changes, x-ray irradiation reduces the electrical resistivity 
by at least 3 orders of magnitude. The triclinic state can be
recovered by thermal annealing. We argue that the transition results from the
x-ray-induced disorder in the Ir$^{4+}$ dimerization pattern, in which 
the orientation of the dimers is preserved, but the translational long-range
order is destroyed.

We are grateful to L. Berman, C. H. Chen, J. P. Hill,
and A. J. Millis for important discussions.
This work was supported by the NSF under grants No. DMR-0103858,
and DMR-0093143, and by A. P. Sloan Foundation (VK).


\begin{references}
\bibitem{Dir} And the symmetry-related [1,-1,0] direction. All the Ir$^{4+}$
dimers are contained in one plain.
\end{references}
\end{document}